\documentclass[preprintnumbers,amsmath,amssymb,floatfix,twocolumn,showpacs,prx,superscriptaddress]{revtex4-1}
\usepackage{etoolbox}
\usepackage{mathrsfs}
\usepackage{dcolumn}
\usepackage{bm}
\usepackage{mathrsfs}
\usepackage{amsthm}
\usepackage{bigints}
\usepackage{float}
\usepackage{color}
\usepackage{graphicx}
\usepackage{subcaption}
\usepackage{multirow}
\newcommand{\beqn}{\begin{eqnarray}}
\newcommand{\eeqn}{\end{eqnarray}}

\newcommand{\ee}{\epsilon}

\begin{document}
\title{Lattice Monte Carlo for Quantum Hall States on a Torus}

\author{Jie Wang}
\affiliation{Department of Physics, Princeton University, Princeton NJ 08544, USA}
\author{Scott D. Geraedts}
\affiliation{Department of Electrical Engineering, Princeton University, Princeton NJ 08544, USA}
\affiliation{Department of Physics, Princeton University, Princeton NJ 08544, USA}
\author{E. H. Rezayi}
\affiliation{Department of Physics, California State University, Los Angeles, CA 90032, USA}
\author{F. D. M. Haldane}
\affiliation{Department of Physics, Princeton University, Princeton NJ 08544, USA}

\begin{abstract}
Monte Carlo is one of the most useful methods to study the quantum Hall problems. In this paper, we introduce a fast lattice Monte Carlo method based on a mathematically exact reformulation of the torus quantum Hall problems from continuum to lattice. We first apply this new technique to study the Berry phase of transporting composite fermions along different closed paths enclosing or not enclosing the Fermi surface center in the half filled Landau level problem. The Monte Carlo result agrees with the phase structure we found on small systems and confirms it on much larger sizes. Several other quantities including the Coulomb energy in different Landau levels, structure factor, particle-hole symmetry are computed and discussed for various model states. In the end, based on certain knowledge of structure factor, we introduce a algorithm by which the lattice Monte Carlo efficiency is further boosted by several orders.
\end{abstract}

\maketitle
\tableofcontents
\section{Introduction}
Numerical Monte Carlo studies of quantum Hall model wavefunctions have long been an important tool in understanding quantum Hall physics. Quantities such as ground state energies, quasiparticle gaps, density-density correlation functions (structure factors), quasiparticle statistics and more have all been calculated using these methods
\cite{PrangeGirvin,laughlin,zhu1993,nonabelian,baraban2009,ciftja2011,biddle2013}. 
Like many numerical methods, these Monte Carlo studies are limited in the system sizes they can access, and methods to increase these system sizes can allow for new measurements and lead to new physical insights.

The torus geometry has been one of the must useful platforms for studying homogeneous Fractional Quantum Hall (FQH) states\cite{haldanetorus1,haldanetorus2}. The translation group of charged particle in a magnetic field on a torus has a rich structure which allows for numerical improvement and deep understandings. Another area of quantum Hall physics of recent interest is a development of a geometric picture in terms of guiding center coordinates\cite{Haldanegeometry}. 
Combining these concepts as recently led one of us \cite{haldanelattice} to show that instead of the continuous wavefunction formalism, a rigorous finite lattice representation can be built on torus, and is applicable for all homogeneous FQH states. 

In this work we show how this lattice representation can be used to significantly speed up Monte Carlo calculations on a torus.
We begin in Section~\ref{MC_calculation_section} with a pedagogical review of the guiding center physics and the translation symmetry on torus, and finally introduce the lattice Monte Carlo method. In the subsequent sections we provide some examples of calculations which can be performed using this new Monte Carlo method. In Section~\ref{sec:phases}, we compute Berry phases for quasiparticles in the Laughlin state, as well as the Berry phase acquired when moving composite fermions around the Fermi sea in a composite Fermi liquid (CFL) state. In Section~\ref{sec:structure}, we compute structure factors for various quantum Hall states at very large sizes, and introduce a Brillouin zone truncation method to significantly improve the lattice Monte Carlo efficiency. Finally in Section~\ref{sec:particlehole}, we show how the Monte Carlo method can be used to evaluate the particle-hole symmetry of wavefunctions in a first-quantized basis.

\section{Lattice Monte Carlo Method}
\label{MC_calculation_section}
In this section, we will introduce the basic notations, and provide a brief review of the guiding center physics and translational symmetry on torus. Both of them played an important role in the development of the ``Lattice representation'' \cite{haldanelattice}.

\subsection{Review of Guiding Center Physics}
A generic quantum hall problem is formed by a 2D electron gas (2DEG) in a high magnetic field. The Hamiltonian that describes this system contains a kinetic term $H_0$ and an interaction term $V$,
\begin{equation}
H = H_0 + \sum_{i < j} V(\bm{r}_i - \bm{r}_j),\quad H_0 = \sum_i\epsilon(\bm{\pi}_i)\label{Hamiltonian}
\end{equation}
where $\epsilon(\bm{p})$ is the single body dispersion and  $\bm{\pi}_i = \bm{p}_i - e\bm{A}(\bm{r}_i)$ is the gauge invariant dynamical momentum.
This momentum satisfies $[\pi_{i,a}, \pi_{j,b}] = i\delta_{i,j}\ee_{ab}l_B^{-2}$ where $l_B=\sqrt{eB/\hbar}$ is the magnetic length and $\ee_{ab}$ is the 2D anti-symmetric symbol (which is odd under time reversal and particle hole conjugation). 
Here the subscripts $i,j$ label different electrons while $a,b$ label directions. The electron's cyclotron motion under this convention is clock-wise, and the ``magnetic area" occupied by one flux quanta is $2\pi l_B^2$. 

The electron's position $\{\bm r_i\}$ can be reorganized to two independent sets, 
\beqn 
r_i^a = R_i^a + \bar R_i^a
\label{LOr_GCr}
\eeqn where $\bar R_i^a\equiv -l_B^2\ee^{ab}\pi_{i,b}$ describes the electrons orbital motion and $R_i^a$ is the guiding center coordinate which is the center of classical cyclotron motion. The algebras for these new coordinates are $[\bar R_i^a, \bar R_j^b] = il_B^2\ee^{ab}\delta_{i,j}$, $[R_i^a, R_j^b] = -il_B^2\ee^{ab}\delta_{i,j}$ and $[\bar R_i^a, R_j^b] = 0$. The kinetic part of the Hamiltonian, $H_0$, produces Landau levels after quantization \cite{Haldanegeometry,haldaneyuiqh}. In the limit when Landau level energy splitting is much larger than the interaction energy, the wavefunction could be written as an \emph{un-entangled product} of the Landau orbit part and the guiding center part
\beqn
|\psi_n\rangle = |\psi^{LO}_n\rangle\otimes|\psi^{GC}\rangle.\label{LOwf_GCwf}
\eeqn
Here $|\psi^{LO}_n\rangle$ is the Landau orbit part, with $n$ indicating that the system is in the $n_{th}$ Landau level. $|\psi^{GC}\rangle$ is the guiding center part of the wavefunction. The Landau level part of wavefunction can be projected out, leaving the problem essentially a degenerate perturbation problem within a specific Landau level described by a set of non-commutative interacting guiding center coordinates $R_i^a$ \cite{Haldanegeometry},
\beqn
H = \sum_{i < j}V(\bm R_i-\bm R_j). \label{gchamiltonian}
\eeqn

In the following we will work on a torus with primary translations $\bm L_1$ and $\bm L_2$, which contains flux $2\pi N_{\phi}=|\bm L_1\times \bm L_2|$. Model wavefunctions on a torus contain an implicit ``complex structure'', which describes a mapping between the torus and the complex plane: $z\equiv w_ax^a$. A complex structure is defined by a unimodular (unit determinant) Euclidean signature metric through $g_{ab} = w_a^*w_b + w_aw_b^*$ and $i\epsilon_{ab} = w_a^*w_b - w_aw_b^*$. The complex lattice is then $\mathbb{L}\equiv\{mL_1 + nL_2\}$, $L_i = w_a\bm{L}_i^a$, and the quantization condition translates to $L_1^* L_2 - L_1L_2^* = 2\pi i N_{\phi}$.

The symmetry group on a torus in a magnetic field is the magnetic translation group, whose group elements are $t(\bm d)\equiv e^{i\bm d\times \bm R}$, where $\bm d$ is a vector in real space. The $t(\bm d)$ satisfies the Heisenberg algebra
\beqn
t(\bm d + \bm d') = t(\bm d)t(\bm d')e^{\frac{i}{2}\bm d\times \bm d'}.\label{Heisenberg}
\eeqn
A periodic translation must leave the wavefunction invariant up to a phase:
\beqn
t(L)\psi(z) = \eta_L^{N_{\phi}}e^{i\phi_L}\psi(z), 
\label{eq:bc}
\eeqn
where $\eta_L = 1$ if $\frac 12 L\in\mathbb{L}$, $\eta_L=-1$ otherwise. The phase $\phi_L$ is usually called the boundary condition. Translating the wavefunction by less than a lattice vector must not change the boundary condition and this forces $d\equiv w_a\bm d^a$ to be quantized with discrete values $mL_1/N_{\phi} + nL_2/N_{\phi}$.

Since wavefunctions on a torus will need to be quasiperiodic, they are naturally expressed in terms of various elliptic functions. In this work the elliptic functions we will use are called `modified Weierstrass sigma functions' $\sigma(z)$ [its definition and one numerical convergent formula is given in Appendix A], which we then multiply by a Gaussian. These functions, which we will call $f(z) = \sigma(z)e^{-\frac{1}{2N_{\phi}}zz^*}$, are building blocks of our model wavefunctions in Section~\ref{sec:phases}.


We need to point out that we are not limiting ourselves in the lowest Landau level by using the holomorphic wavefunctions. Holomorphic functions are just representation of guiding center algebras and are generic to any Landau level.

\subsection{Lattice Monte Carlo Method}
A number of useful calculations (i.e. overlap, operator expectation value) can be made by integrating the positions of all electrons in a model wavefunction. In this short section, we will review the definition and derivation of the ``lattice representation'' \cite{haldanelattice}, and then describe how these calculations can be performed using the Metropolis-Hastings algorithm \cite{Metropolis,Hastings}. We'll focus on two-body operators since they appear in useful quantities such as energy, structure factor et.al.

The key advantage of our method lies in the fact that continuous integration can be replaced with lattice summation on torus in an exact way: if we are interested in knowing the mean value of a translational invariant two-body operator $\sum_{i<j}O(x_i-x_j)$, averaged by states $|\psi_{n,1}\rangle$ and $|\psi_{n,2}\rangle$ in the $n_{th}$ Landau level, which by definition is given by continuous integration,
\beqn
&&\langle\psi_{n,1}|\hat O|\psi_{n,2}\rangle \nonumber\\
&\equiv&\prod_k^{N_e}\int_{\Omega}d^2x_k\ \psi_{n,1}^*(\{x\})\psi_{n,2}(\{x\})\sum_{i<j}O(x_i-x_j).\nonumber
\eeqn
In fact, such calculation can be replaced by a lattice summation for operator $\sum_{i<j}O^{Lat}(x_i-x_j)$, which we called as the ``lattice representation'' of $\hat O$,
\begin{eqnarray}
\langle\psi_{n,1}|\hat O|\psi_{n,2}\rangle_{} &=& C \langle\psi_{0,1}| \hat O^{Lat} |\psi_{0,2}\rangle_{Lat}. \label{latticesum_op}
\end{eqnarray}
where the symbol $\langle|...|\rangle_{Lat}$ means lattice summation,
\beqn
&&\langle\psi_{0,1}| \hat O^{Lat} |\psi_{0,2}\rangle_{Lat}\nonumber\\
&\equiv& \prod_k^{N_e}\sum'_{x_k}\ \psi_{0,1}^*(\{x\})\psi_{0,2}(\{x\})\sum_{i<j}O^{Lat}(x_i-x_j).\nonumber
\eeqn
In the above, $\sum'_{x}$ means summing over the $N_{\phi}\times N_{\phi}$ evenly spaced lattice. 
The constant $C$ is fixed once the $N_{\phi}\times N_{\phi}$ lattice is chosen, and it is not important since it is always canceled out by wavefunction normalization factors. The $\hat O^{Lat}$ dependents on the lattice choice, whose expression will be given soon. Note that when $O(x)$ is the identity operator, (\ref{latticesum_op}) means that wavefunction overlap can be calculated from lattice sum.

In Eq.~(\ref{latticesum_op}), we wrote states on the right side of the equation with Landau level index $n=0$. By doing this, we mean that we can solve the physical problem in an arbitrary Landau level by using the lowest Landau level lattice representation. 

The translation group plays the central role in the derivation of the lattice representation. To see this, we start by finding the effective interaction potential for guiding centers. First, do a Fourier expansion for $\hat O$, yielding,
\beqn
\langle x_i|\hat O|x_j\rangle \equiv
O(x_i - x_j) = \frac{1}{2\pi N_{\phi}} \sum_q O(q)e^{iq(x_i-x_j)}\nonumber
\eeqn
where unprimed sum sums all discrete $q$ allowed by boundary condition.
Now we split up the coordinate and wavefunction into `Landau orbit' and `guiding center' parts using Eqs.~(\ref{LOr_GCr}) and (\ref{LOwf_GCwf}).   
This allows us to write it as,
\beqn
&&\langle\psi_{n,1}|\hat O|\psi_{n,2}\rangle = \nonumber\\
&&\frac{1}{2\pi N_{\phi}}\sum_q\sum_{i<j}O(ql_B)f_n^2(ql_B)\langle\psi_1^{GC}|e^{iq(R_{i}-R_{j})}|\psi_2^{GC}\rangle. \label{transla}
\eeqn
where $f_n(ql_B)$ is the Landau level `form factor':
\beqn
f_n(ql_B) \equiv \langle \psi^{LO}_n| e^{iq\bar R} |\psi^{LO}_n\rangle = L_n(\frac12q^2l_B^2)e^{-\frac{1}{4}q^2l_B^2}.
\label{form_factor}
\eeqn
A key observation is that $e^{iqR}$ in Eq.~(\ref{transla}) is nothing but the magnetic translation operators, which satisfy the Heisenberg algebra Eq.~(\ref{Heisenberg}). Note that the periodic translation ($q\in\mathbb{L}$) leaves the state invariant up to a phase factor Eq.~(\ref{eq:bc}). We thus have broken up the sum over $q$ into a sum over the first Brillouin zone [indicated by the prime on the sum], and the sum over the rest of $q$ space included in $O^{GC}(ql_B)$,
\beqn
&&\langle\psi_{n,1}|\hat O|\psi_{n,2}\rangle = \nonumber\\
&&\frac{1}{2\pi N_{\phi}}\sum'_q\sum_{i<j}O^{GC}(ql_B)\langle\psi_1^{GC}|e^{iq(R_{i}-R_{j})}|\psi_2^{GC}\rangle\nonumber\\\label{latop}
\eeqn
where $O^{GC}(ql_B)$ is the effective interaction defined in the first Brillouin zone acting on the guiding centers,
\beqn
O^{GC}(ql_B) &=&[O(ql_B)f_n^2(ql_B)]_c \nonumber\\
&\equiv&\sum_{q'} O(ql_B+q'N_{\phi}l_B)f_n^2(ql_B+q'N_{\phi}l_B)\nonumber
\eeqn
%
To find out the expression for the lattice representation $O^{Lat}$, we need to look at single-body operators. We do not attempt to include more derivation details here, but leave them in the Appendix B. We refer the readers to \cite{haldanelattice} for more details. Its expression, which is the central result of this section, is:
\beqn
O^{Lat}(x) = \frac{1}{2\pi N_{\phi}} \sum'_{q}\frac{O^{GC}(ql_B)}{|[f_0(ql_B)]_{N_{\phi}}|^2}e^{iqx}.\label{latticesum2}
\eeqn
The $[...]_{N_{\phi}}$ around the form factor is a notion of compactification which indicates that it, like $O^{GC}$, is summed over all Brillouin zones (definition of $[...]_{N_{\phi}}$ is Eq.~(\ref{f0nphi}) in Appendix B).

At this stage, we make some comments on the result. The emergence of the $q-$space Brillouin zone in Eq.~(\ref{latop}) is purely a consequence of the translation group (\ref{Heisenberg}), and this indeed implies the real space lattice structure: states and operators can be formulated in an exact way on lattice. Besides, since we worked out the whole problem in the guiding center space, the lattice representation is generic to any Landau level; the lowest Landau level wavefunction in Eq.~(\ref{latticesum_op}) is not special, but serves just as a technique device to solve the problem in a generic Landau level. Furthermore, in some cases when the two body operator $O(q)$ is divergent if it's put on the infinite plane, their lattice representations are convergent. Seen from Eq.~(\ref{latticesum2}), the numerator and denominator are regularized by Gaussian factor first and are compacified separately, making the potential convergent. This is not surprising, since lattice provides a nature regularization. We will meet them when working on the high Landau level Coulomb energy and pair-amplitude.

We thus finished the discussion on the lattice representation. To adopt the Metropolis algorithm, we rewrite expectation value as:
\begin{eqnarray}
&&\frac{\langle\psi_1|\hat O|\psi_2\rangle}{\sqrt{\langle\psi_1|\psi_1\rangle\langle\psi_2|\psi_2\rangle}} \label{MCdef} \\
&=&\frac{[\sum' |\psi_1(x)|^2\cdot O^{Lat}(x)\cdot\psi_2(x)/\psi_1(x)]/\sum'|\psi_1(x)|^2}{\sqrt{[\sum' |\psi_1(x)|^2\cdot|\psi_2(x)/\psi_1(x)|^2]/\sum'|\psi_1(x)|^2}}. \nonumber
\end{eqnarray}
We obtained this equation by writing the overlaps as sums over all positions of the coordinates, and then multiplying the numerator and denominator by 
$|\psi_1(x)|^2/[|\psi_1(x)|^2 \sum^\prime_{x^\prime} |\psi_1(x^\prime)|^2]$.
In the above, $\sum'$ sums over $x=\{z_1, ... z_{N_e}\}$ which represents a point in the many body coordinate space. All $z_i$ live on the lattice, therefore $x$ is $N_{\phi}^{2N_e}$ dimensional. Writing the overlap in this way makes it clear that both the numerator and denominator can be computed using a Monte Carlo algorithm with Metropolis weight $|\psi_1|^2$. 

In Table~\ref{table:energy}, we test our Monte Carlo method by computing the Coulomb energy (i.e.\ 
$O(x)\rightarrow V(x)=1/|x|$
) for the Laughlin wavefunction at $\nu=1/3$ in the first few Landau levels [wavefunction is provided in Eq.~(\ref{laughlin})]. 
The tables shows the exact energies and those determined by Monte Carlo, for a few different system sizes. The energy produced by the Monte Carlo does not include the `Madelung energy' (the energy due to an electron's interaction with periodic copies of itself), but this can be calculated analytically\cite{Yoshioka,Bonsall}. The fact that our results agree to several digits (limited only by the statistical error of the Monte Carlo) is a confirmation that our lattice Monte Carlo does give correct results.
For $n>1$ we find very large statistical errors which prevent us from obtaining the energy directly through the Monte Carlo. The cause to this problem and a solution which improves the Monte Carlo efficiency significantly are provided in Section~\ref{sec:structure}.
\begin{table}
\begin{tabular}{|c|c|c|}
\hline
\multicolumn{3}{|c|}{$n=0$} \\
\hline
 $N_e$ & Exact & Monte Carlo  \\
 \hline
$4$ & $-0.414171$ & $-0.414172\pm 0.000001$ \\
$5$ & $-0.412399 $ & $-0.412397\pm 0.000001$ \\
$6$ & $-0.411583 $ & $-0.411585\pm 0.000001$ \\
 \hline\hline
\multicolumn{3}{|c|}{$n=1$} \\
\hline
 $N_e$ & Exact & Monte Carlo  \\
\hline
$4$ & $-0.339105$ & $-0.33907 \pm 0.00005$ \\
$5$ & $-0.334207$ & $-0.33421 \pm 0.00004$ \\
$6$ & $-0.331879$ & $-0.33190 \pm 0.00007$ \\
\hline\hline
\multicolumn{3}{|c|}{$n=2$} \\
\hline
 $N_e$ & Exact & Monte Carlo  \\
\hline
$4$ & $-0.280537$ & $-0.278 \pm 0.004$ \\
$5$ & $-0.278052$ & $-0.280 \pm 0.005$ \\
$6$ & $-0.274849$ & $-0.26 \pm 0.01$ \\
\hline\hline
\multicolumn{3}{|c|}{$n=3$} \\
\hline
 $N_e$ & Exact & Monte Carlo  \\
\hline
$4$ & $-0.257681$ & $-0.26 \pm 0.09$ \\
$5$ & $-0.254155$ & $-0.4 \pm 0.2$ \\
$6$ & $-0.251042$ & $-0.7 \pm 0.9$ \\
\hline
\end{tabular}
\captionsetup{justification=RaggedRight, singlelinecheck=true}
\caption{Comparison of exact and Monte Carlo energies for the Laughlin wavefunction at $\nu=1/3$. The agreement between the two, limited only by statistical error, is a confirmation that our lattice Monte Carlo is correct. For $n>1$ the statistical error is large. The cause and solution of this problem are described in Section~\ref{sec:structure}.}
\label{table:energy}
\end{table} 

\section{Berry Phase}
\label{sec:phases}
In this section we use the Monte Carlo method to calculate the Berry phases acquired when various quasiparticles are moved around a closed path. We will start from an easy case of moving quasi-hole in the Laughlin state. This serves as an example of Berry phase calculation and demonstrating our Monte-Carlo method works. Then we will proceed to the gapless CFL phase. The Berry phase obtained by the composite fermions as they move around the Fermi surface has attracted recent interest due to its relationship with various particle-hole symmetric theories of the CFL\cite{Son,WangSenthil,GeraedtsScience,scottjiehaldane}.

\subsection{Laughlin-Hole Berry Phase}
As a first example we move one quasi-hole in the $\nu=1/q$ [in this section, we'll use $q$ as inverse filling] Laughlin state around an undefected area $A$. Since the quasi-hole is charged and there is a magnetic field passing through the system, the quasi-hole should pick up a Berry phase of $2\pi A/q$. Before doing the Berry phase calculation, let's review the Laughlin and Laughlin-Hole wavefunction on torus.

On the infinite plane, Laughlin's $\nu=1/q$ wavefunction \cite{laughlin} is given by $\prod_{i<j}(z_i-z_j)^qe^{-\frac{1}{2l_B^2}\sum_iz_iz_i^*}$. The torus generalization of it is: \cite{haldanetorus1, haldanetorus2},
\beqn
\Psi(\{\alpha\}) &=& \prod_{i<j}^{N_e}[f(z_i-z_j)]^q\prod_{k=1}^qf(Z - \alpha_k).\label{laughlin}
\eeqn
where $Z = \sum_i^{N_e}z_i$ is the center-of-mass coordinate. The first term of Eq.~(\ref{laughlin}) is the usual Vandermonde factor on a torus, while the second term places $q$ fold center-of-mass zeros at positions $\{\alpha_k\}$. From now on we will enforce periodic boundary conditions by requiring that $\sum_k^q\alpha_k=0\mod L$. The $f(z)$ function is given in the end of the first subsection of Section~\ref{MC_calculation_section}.

Inserting additional $N_h$ fluxes in the $\nu=1/q$ Laughlin wavefunction creates a quasi-hole excitation. The wavefunction with $\{w\}$ representing positions of quasiholes is,
\begin{eqnarray}
&& \Psi(\{\alpha\}, \{w\}) \label{Laughlin-hole}\\
&=& \prod_{i<j}^{N_e}f^q(z_i-z_j)\prod_{i,a}^{N_e,N_h}f(z_i-w_a)\prod_{k=1}^qf(Z + \frac{W}{q} - \alpha_k).\nonumber
\end{eqnarray}

In the following, we will use the Monte Carlo method to calculate this Berry phase $\Phi$. We take the one-hole model wavefunction $N_h=1$, and move it around path $x_0$, $x_1$ ... $x_{n-1}$. At each step, we compute the overlap between the wavefunction with $x=x_n$ and $x_{n+1}$. To compute the Berry phase, we take the product of these overlaps:
\beqn
\langle\psi(x_0)|\psi(x_1)\rangle...\langle\psi(x_{n-1})|\psi(x_n)\rangle = |D| e^{i\Phi}.\nonumber
\eeqn
Since our numerics turns the continuous motion of the quasi-hole into a series of discrete steps, the amplitude $|D|$ will be smaller than one. The system has probability $1-|D|$ jumping to the excited state and scrambling the phase. Therefore it is important to keep the step length $|x_i-x_{i+1}|$ small so that $|D|$ is close to one.

\begin{figure}
    \centering
    \includegraphics[width=0.45\textwidth]{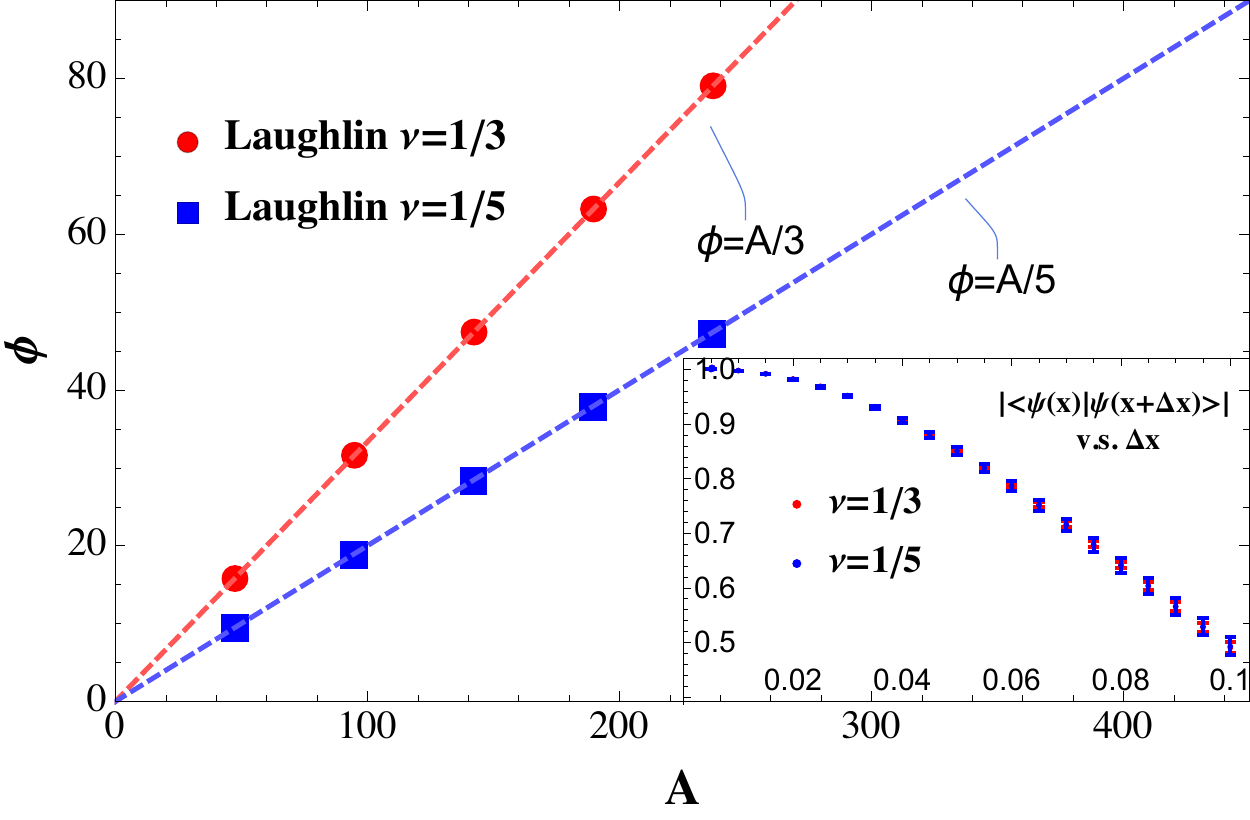}
    \captionsetup{justification=RaggedRight, singlelinecheck=true}
    \caption{ $N_e=50$, $\nu=1/q$. Laughlin-Hole State Berry phase ($\Phi$) v.s. area the loop enclosed ($A$). Blue circle and yellow square are Monte-Carlo data for $q=3$ and $q=5$ respectively. The Monte Carlo error is bound by the line width. This demonstrates $\Phi = A/q$. The inside figure is the overlap $|\langle\psi(0)|\psi(\Delta x)\rangle|$ v.s. $\Delta x$. This allows us to take steplength to be $\Delta x=  0.02$.}\label{laughlinholebp}
\end{figure}

The numerical results for Laughlin $q=3$ and $q=5$ states are represented in (Fig. \ref{laughlinholebp}). We see that our observed values are what we expect them to be. 

Certainly, one can do braiding of holes, or even more exotic anyons in other topological states \cite{nonabelian,baraban2009}. Here we just use Laughlin hole as a trivial example to illustrate the Berry phase calculation.

\subsection{CFL Berry Phase}
The composite-Fermi-liquid state is a gapless state that forms at Landau level filling $\nu=1/q$ when $q$ is even. An emergent surface of composite fermion forms. In this subsection, we will calculate the Berry phase acquired as moving one composite fermion around the Fermi surface.

There are some model wavefunctions proposed for the CFL state, such as $\det_{ij}e^{id_i\cdot R_j}|\Psi_L^{\frac12}\rangle$ \cite{RezayiReadCFL,RezayiHaldaneCFL} where $|\Psi_L^{\frac12}\rangle$ is the boson Laughlin state. Evaluating this wavefunction when projecting to a single Landau level unfortunately requires anti-symmetrization of $N_e!$ terms, and therefore quickly becomes unfeasible for practical calculation when $N_e$ is large. In this work we consider instead the following model wavefunction \cite{JainKamilla,shaoprl}, whose computational complexity is $O(N_e^3)$. 
\begin{equation}
\Psi_{CFL}(\{\alpha\},\{d\}) = \det\tilde M_{ij}\prod_{i<j}^{}f^{q-2}(z_i-z_j)\prod_{k=1}^qf(Z-\alpha_k).\label{CFLwf}
\end{equation}
where $\tilde M_{ij}$ is a $N_e\times N_e$ matrix. The $\det\tilde M_{ij}$ is its determinant,
\begin{equation}
\tilde M_{ij} = e^{\frac{1}{2q}(z_id_j^* - z_i^*d_j)}\prod_{k\neq i}^{N_e}f(z_i-z_k-d_j+\bar d).\nonumber
\end{equation}
In addition to a dependence on the $q$ fold center of mass zeros $\{\alpha\}$, this wavefunction depends on $N_e$ additional parameters $\{d\}$, the dipole moments. Like many quantum Hall wavefunctions, this wavefunction surrounds each electron with a `correlation hole': a region of depleted charge. In this wavefunction, the center of the correlation hole is displaced from the electron by $d_j$. In the magnetic field, dipolar electron always moves perpendicular to its dipole direction, therefore the composite fermion's momentum is $k_a = l_B^{-2}\ee_{ab}ed^b$.

Requiring that all electrons see the same boundary conditions sets some constraints on the $\{d\}$. 
The total zeros seen by the $i^{th}$ particle add up to $\alpha - d + N_ed_{P(i)}$. Since all electron must satisfy the same boundary condition, $d_i$ must take $\{m\frac{L_1}{N_e} + n\frac{L_2}{N_e}\}$ values.
Of course this quantization makes sense if we remember that the $l_B^{-2}\ee_{ab}ed^b$ represent composite fermion momentum, and momentum is quantized on torus. 



From the numerical work we have done on small system sizes \cite{scottjiehaldane}, we know that the model wavefunction is very close to the Coulomb ground state when the dipoles are clustered, and becomes less close when more dipoles are excited out of Fermi sea. 
We first need to define what it means to take a composite fermion around the Fermi sea. In this work we consider a set of states obtained from dipole moments which form a compact Fermi sea, plus one additional dipole moment. We move this dipole moment on a path which encloses the Fermi sea. Alternatively we can remove a dipole moment which corresponds to taking a composite hole around the Fermi sea.
Because the many body momentum $K = \sum_i^{N_e}d_i$ [which is the eigen-value of many-body translation operator], these states defines a path in the momentum space.

Since our system has translation invariance, states with different momentum are generally orthogonal $\langle\psi(K_1)|\psi(K_2)\rangle = 0$ if $K_1\neq K_2$. We must insert an operator that makes this overlap non-vanishing. The natural choice of this operator is the guiding center density operator $\rho(\bm d) = \sum_i^{N_e}t_i(\bm d)$ which satisfies the GMP algebra $[\rho(\bm d_1), \rho(\bm d_2)] = 2i\sin\frac{\bm d_1\times \bm d_2}{2l_B^2}\rho(\bm d_1+\bm d_2)$. We thus defined the many-body $K$ space Berry phase, which is a generalization of the single body Brillouin zone Berry phase, as follows,
\beqn
|D|e^{i\Phi}=Tr(\bm{\Gamma}_{12}\bm{\Gamma}_{23}...\bm{\Gamma_{N,1}}).
\eeqn
where for each step $(\bm{\Gamma}_{1,2})_{\alpha,\beta} \equiv \langle\psi_{1\alpha}|\rho(\Delta K_{12})|\psi_{2\beta}\rangle$ with $\Delta K_{12}$ takes value in the first Brillouin zone and $\Delta K_{12} = K_1-K_2 \mod L$ . Here the $\alpha,\beta = 0,1$, labels the two-fold topological ground states. The off diagonal elements of $\Gamma_{1,2}$ are small since they involve transition between different topological sectors.

In addition to the phase we are interested in, the phase $\Phi$ contains a contribution from the density operator. 
From \cite{scottjiehaldane}, we have found that this phase is determined by the direction the composite fermion moves around the Fermi sea. The total phase is given by:
\beqn
e^{i\Phi} = (i)^{N_+ - N_-} (-1)^{\eta}
\label{berryphase}
\eeqn
In the above formula, $(i)^{N_+ - N_-}$ is a path-dependent phase, and $(-1)^{\eta}$ is the $Z_2$ part. $N_+$ ($N_-$) is the number of anti-clock (clock) wise steps, defined relative to the center of Fermi sea. Note that steps normal to the Fermi sea are not included, since they always have zero amplitude. The $\eta\in\mathbb{Z}$ is the winding number, counting how many times the total path enclose the center of the Fermi sea.

\begin{figure}
    \centering
    \includegraphics[width=0.4\textwidth]{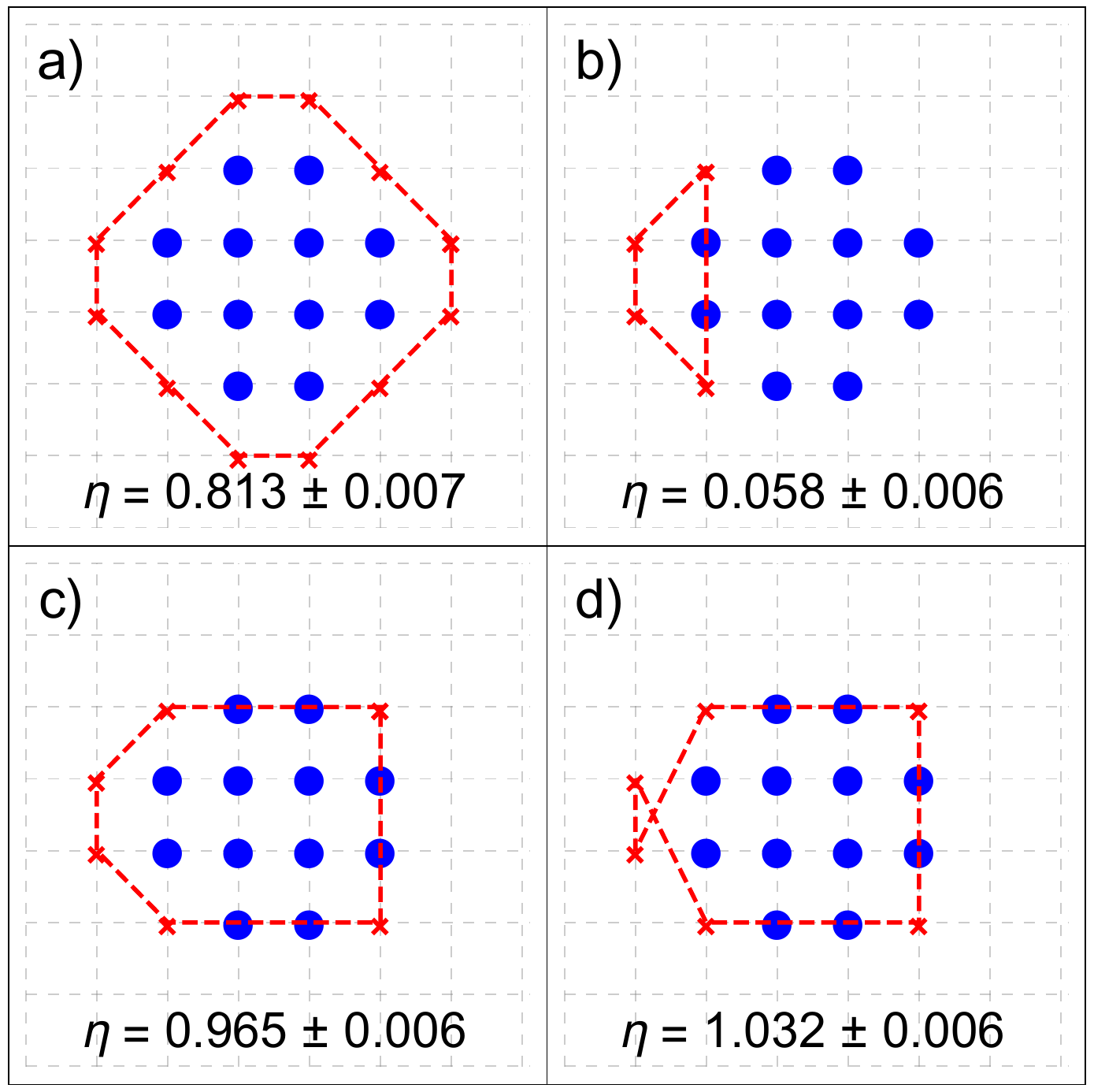}
    \captionsetup{justification=RaggedRight, singlelinecheck=true}
\caption{ $N_e=13$ CFL Berry phase. Cross marks represent the `composite fermions' we are moving. This is a consistency check with the same calculation  done in \cite{scottjiehaldane} (but using a different numerical approach). This data can be interpreted through Eq.(\ref{berryphase}), which shows that in addition to a $\mathbb{Z}_2$ piece there is a piece depending on the direction of motion around the Fermi surface. When this is accounted for we find a ``-1'' from the $\mathbb{Z}_2$ part whenever the composite fermion encloses the Fermi sea. }\label{cflbp13}
\end{figure}

\begin{figure}
    \centering
    \includegraphics[width=0.45\textwidth]{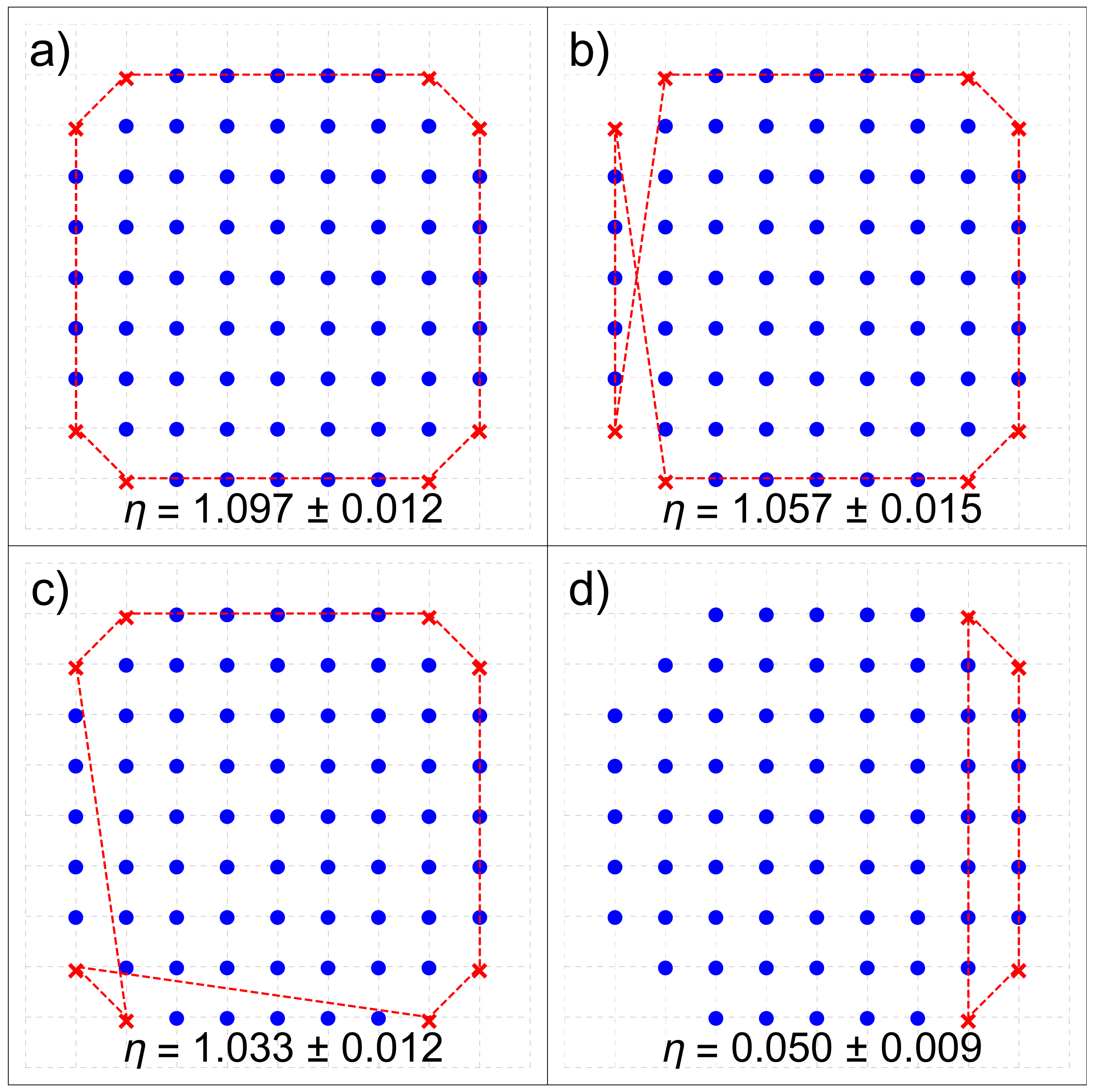}
    \captionsetup{justification=RaggedRight, singlelinecheck=true}
    \caption{$N_e=69$ CFL Berry Phase. Cross marks represent the `composite holes' we are moving. The results are again consistent with Eq.~(\ref{berryphase})}\label{cflbp69}
\end{figure}

The Monte Carlo enables us to look at the Berry phase on much larger sizes up to $N_e=69$, and let us to check the Berry phase in a more convincing way. The following (Fig. \ref{cflbp13}) is done for $N_e=13$, and (Fig. \ref{cflbp69}) is for $N_e=69$. The results agree with Eq.~(\ref{berryphase}), confirming that a $\mathbb{Z}_2$ phase is indeed obtained when composite fermions encircle the origin. The $\eta$ computed from Monte Carlo is close but not exactly $\pm 1$ because the model wavefunction is not exactly particle-hole symmetric.

\section{Structure Factor and Pair Amplitude}
\label{sec:structure}
Another application of the Monte Carlo technique is the (static) guiding center structure factor $S(q)$ which plays an important role in the FQH.

In the ``single-mode approximation'' first introduced by Feynman in superfluid Helium-4 \cite{Feynman} and then adopted by Girvin, MacDonald and Platzman in FQH\cite{gmpl,gmpb}, the structure factor provides a variational upper bound of the neutral excitations. In particular, the $|q|\rightarrow0$ behavior of $S(q)$ is closely related to the collective modes in the system, and is a criteria whether the system is gapped or not at long wavelength. For example, in superfluid Helium-4, $S(q)\sim |q|^2$, corresponds to the gapless phonon mode, while in FQH $S(q)\sim |q|^4$ corresponds to the gapped graviton mode \cite{Haldanegeometry}. For Laughlin wavefunction, the $4^{th}$ and $6^{th}$ order expansion coefficient of $S(q)$ are predicted in \cite{sixordersq}. The larger sizes accessible using our Monte Carlo method allow us to test these predictions.

Additionally, for the gapless CFL state, the peak in structure factor can used to identify the composite fermion Fermi surface, and identify its symmetry properties\cite{GeraedtsScience}. We can observe this physics in our Monte Carlo data. Lastly, from the structure factor, we found a method to greatly improve the Monte Carlo efficiency.

\subsection{Structure Factor}
The guiding center (static) structure factor by definition is the density-density correlation function,
\beqn
S(q) \equiv \frac{1}{2N_{\phi}} \langle\psi^{GC}| \{ \delta\rho(\bm q), \delta \rho(-\bm q)\} |\psi^{GC}\rangle. \label{GC_structure_factor}
\eeqn
where $\delta \rho(\bm q)$ is the fluctuation of density operator $\rho(\bm q)$ relative to the background $\langle \rho(\bm q) \rangle/N_{\phi} = 2\pi l_B^2 \nu \delta^2(\bm ql_B)$,
\beqn
\rho(\bm q) &=& \sum_i^{N_e} e^{iq_aR_i^a}, \nonumber\\
\delta \rho(\bm q) &=& \rho(\bm q) - \langle \rho(\bm q) \rangle. 
\eeqn
Note that both $\delta\rho(\bm q)$ and $\rho(\bm q)$ satisfy the GMP algebra.

\begin{figure}
    \centering
    \includegraphics[width=0.8\linewidth]{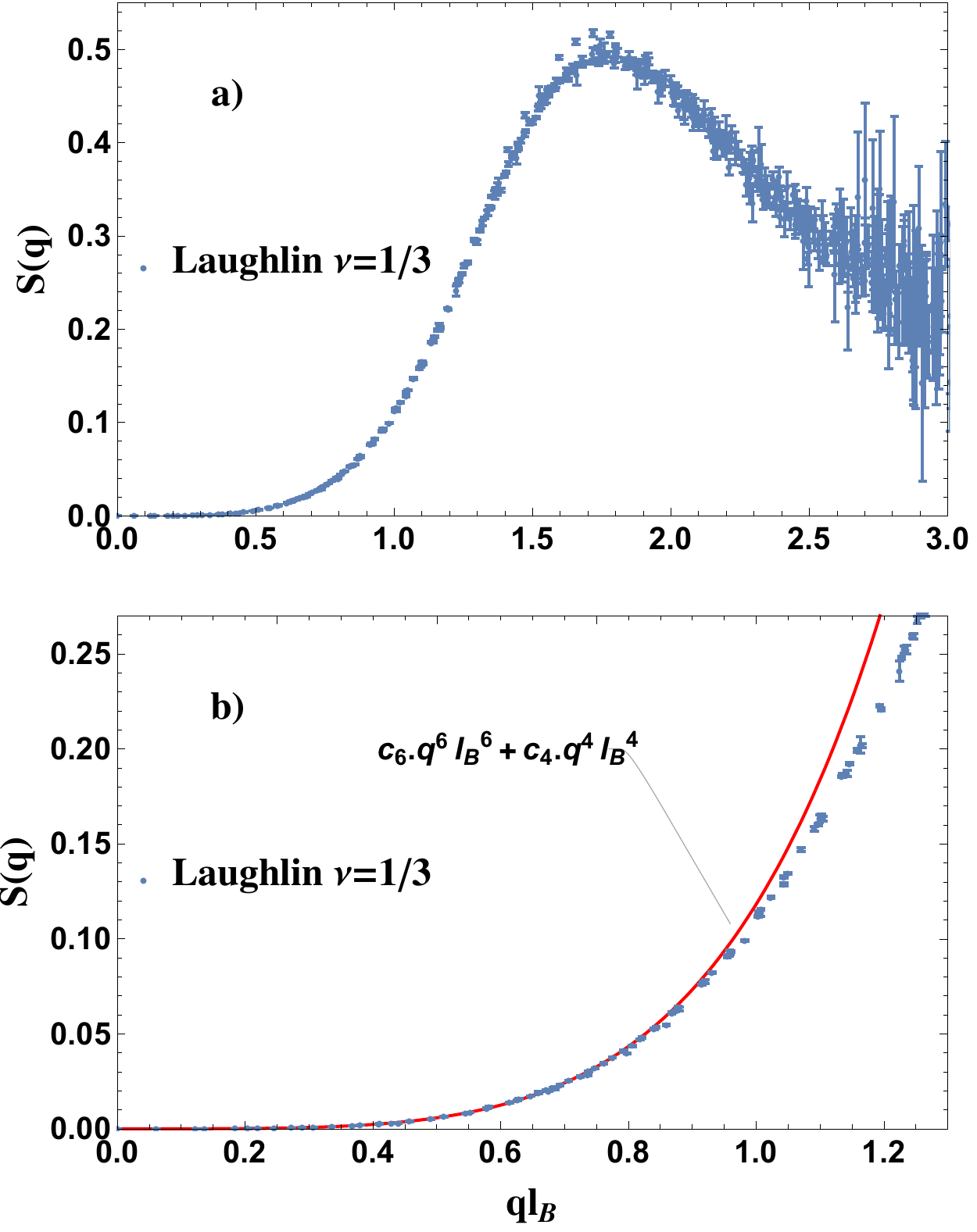}\label{fig:lausq}
    \captionsetup{justification=RaggedRight, singlelinecheck=true}
    \caption{The guiding center structure factor for $N_e=50$ electron in Laughlin $\nu=1/3$ state. The subfigure $a)$ is its plot together with error bar. The Gaussian function $e^{-\frac{1}{4}q^2l_B^2}$ limits us to see $S(q)$ only within a window. In subfigure $b)$, we check long-wavelength expansion ($c_4$ and $c_6$) in (\ref{exactexpansion}) by comparing the Monte Carlo data and it, where $c_4, c_6$ given by (\ref{exactcurve6}) and all other $c_i=0$. It can be seen that the long-wavelength behavior of $S(q)$ is correctly described by (\ref{exactcurve6}).}\label{lausq}
\end{figure}

Several properties of $S(q)$ are worth to be mentioned \cite{haldaneselfdual}. First, the large $|q|$ asymptotic value is determined by filling $S(\infty) = \nu(1+\xi \nu)$, where $\xi=-1$ if the underlying particles are fermions, $=1$ is bosons. Second, $S(q)$ is self-dual under Fourier transformation,
\beqn
S(\bm q)-S(\infty) = \xi\int \frac{d^2\bm q'l_B^2}{2\pi}e^{i\bm q\times \bm q'l_B^2}(S(\bm q') - S(\infty)). \label{self-dual}\nonumber\\
\eeqn
Third, the coefficients of the small $q$ expansion
\beqn
S(q) = c_2 |q|^2 + c_4 |q|^4 + c_6 |q|^6 + ..., \label{exactexpansion}
\eeqn
contain useful information.
For a gapless system $c_2\neq0$, while for a gapped system $c_2=0$. For a Laughlin $\nu=1/q$ state, $c_2=0$, and predictions exist for $c_4$ and $c_6$\cite{sixordersq}:
\beqn
c_4 &=& \frac{\nu|s|}{4}, \nonumber\\
c_6 &=& \frac{\nu|s|}{8}(s - \frac{c-\nu}{12}\frac{1}{\nu s}). \label{exactcurve6}
\eeqn
where $s = -\frac{1}{2}(q-1)$ is the guiding center spin \cite{Haldanegeometry}, $c$ is the central charge. Our Monte Carlo method allows us to test these predictions (Fig. \ref{lausq}).

Another way to write Eq.~(\ref{GC_structure_factor}) is as follows:
\beqn
&&S(q)= \nonumber\\
&&\frac{1}{N_{\phi}} \sum_{i,j}^{N_e} \langle \psi^{GC}| e^{iq(R_i - R_j)} |\psi^{GC} \rangle 
- \frac{1}{N_{\phi}}\langle\rho(q)\rangle\langle\rho(-q)\rangle.\nonumber
\eeqn
Writing $S(q)$ in this way reveals a challenge when computing it with our Monte Carlo method, which computes expectation values relative to the real-space coordinates $r$ and Schr\"odinger wavefunctions, rather than the guiding center versions, see Eq.~(\ref{LOr_GCr}-\ref{LOwf_GCwf}).
What our Monte Carlo calculates is the ``full structure factor'' (per flux), defined as:
\beqn
&&S^{full}(q)= \nonumber\\
&& \frac{1}{N_{\phi}} \sum_{i,j}^{N_e} \langle \psi_0| e^{iq(r_i - r_j)} |\psi_0 \rangle_{Lat} 
- \frac{1}{N_{\phi}}\langle\rho(q)\rangle\langle\rho(-q)\rangle.\label{structure3}
\eeqn
We can relate these two quantities by using the form factor defined in Eq.~(\ref{form_factor}) and Eq.~(\ref{twobodylat}) to simplify $S^{full}(q)$.
This shows that $S^{full}(q)$ is related to the guiding center structure factor $S(q)$ via,
\beqn
S^{full}(q) -\nu= |[f_0(q)]_{N_{\phi}}|^2\cdot [S(q)-\nu]
\label{structure_relation}
\eeqn
The $\nu$ in the above equation comes from the terms in the sum where $i=j$. 
Because of the Gaussian function $f_0(q) = e^{-\frac{1}{4}q^2l_B^2}$, the Monte Carlo error in $S^{}(q)$ is amplified greatly when $|q|$ is large. This limits us to see the $S(q)$ within a window. For Laughlin, $|q_{max}| \approx 3l_B^{-1}$ (Fig. \ref{lausq}).

\begin{figure}[h]
    \centering
    \includegraphics[width=0.45\textwidth]{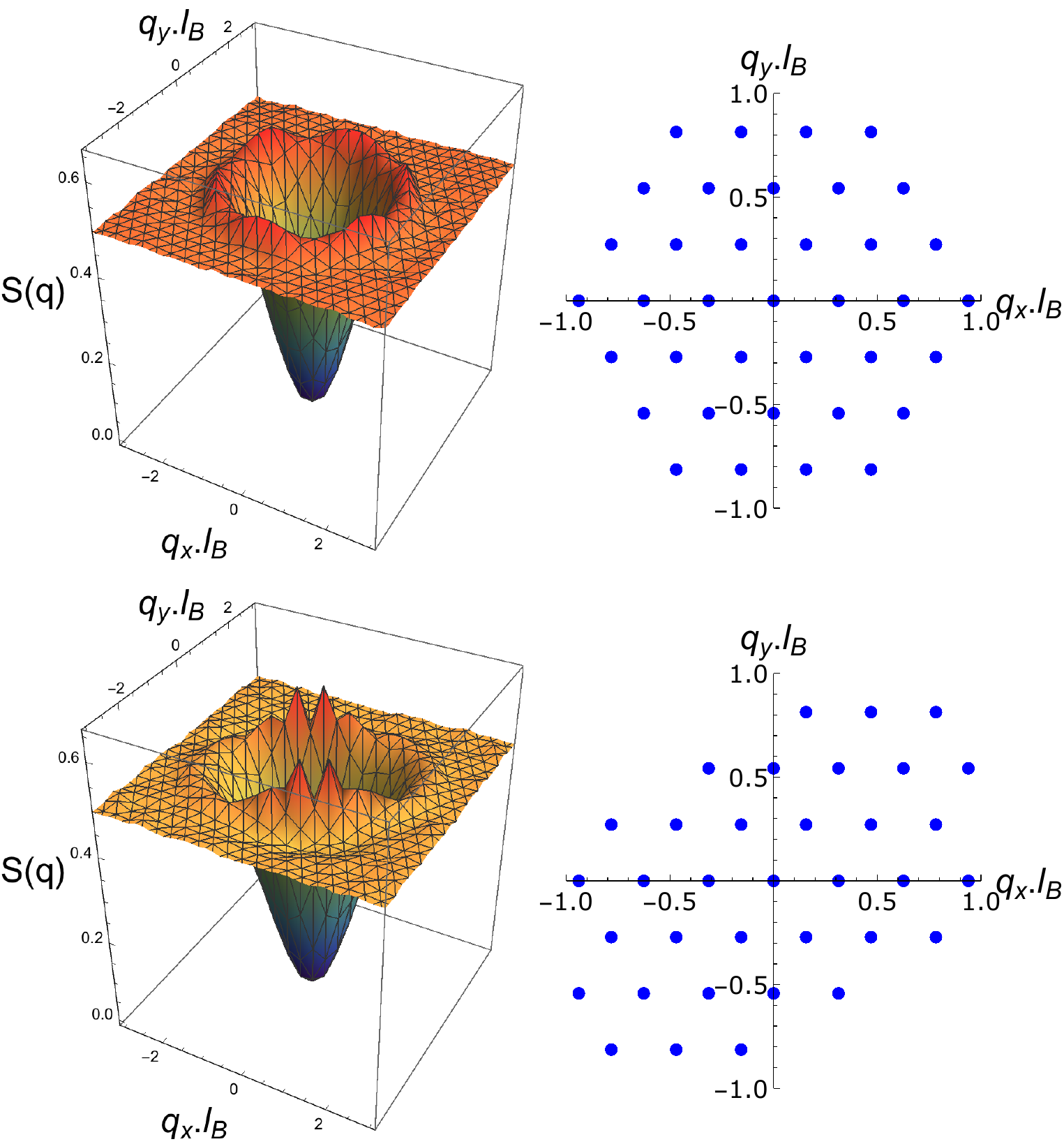}
    \caption{CFL $S(q)$ and dipole configuration. The peak of the structure factor and composite fermion Fermi surface have the same shape. The radius of the former is twice of the latter's.}\label{cflsq}
\end{figure}

For the CFL states, the shape of the Fermi surface can be read from the peak of structure factors \cite{GeraedtsScience}. And the radius of the latter should be twice as large as that of the former. Here we plot the structure factor for model wavefunctions with different dipole moment configurations for $N_e=37$ electrons (Fig. \ref{cflsq}).

\subsection{Improved Monte Carlo Algorithm}
In Section~\ref{MC_calculation_section}, we showed how to calculate any two-body expectation value $O(x_i-x_j)$, in any Landau level, and to demonstrate our method we computed the Coulomb energy of the Laughlin state in the first two Landau levels. However we found that for higher Landau levels ($n>1$), our method was subject to large Monte Carlo errors. In this section we will use our insights about the structure factor to understand and ameliorate these errors. The algorithm discussed in this section applies to other translational invariant two body interactions, like pair-amplitude. We will borrow the notions from Section~\ref{MC_calculation_section}.

The first step in this process is to find out the effective potential acting on the guiding centers,
\beqn
\langle\hat O\rangle &=& \frac{1}{2\pi N_{\phi}}\sum'_q \sum_{i<j} O^{GC}(ql_B) \langle e^{iq(R_i-R_j)}\rangle\nonumber\\
&=& \frac{1}{4\pi N_{\phi}}\sum'_{q} O^{GC}(ql_B) \cdot [S(ql_B) - \nu].\label{Oxp2}
\eeqn
In the problem of high Landau level Coulomb energy ($n$ is the Landau level index),
\beqn
O^{GC}(ql_B) &=& [O(ql_B)f_n^2(ql_B)]_c,\nonumber\\
O(ql_B) &=& \frac{2\pi}{|ql_B|},\quad |ql_B|\neq 0.\label{highcoulombogc}
\eeqn
Equation Eq.~(\ref{Oxp2}) tells us that, at least in principle, the guiding center structure factor allows us to calculate any expectation value. However, we found in the previous section that $S(q)$ determined from our Monte Carlo procedure has very large errors as at large $|q|$. The reason these errors don't completely ruin our calculation is that Eq.~(\ref{Oxp2}) also contains a form factor $f_n(q)$, which decays to $0$ exponentially at large $|q|$, thus suppressing the errors. Unfortunately, the decay of $f_n(q)$ gets weaker as the Landau level index $n$ is increased. This is why we we had difficulty  calculating Coulomb energies for $n>2$ in Section~\ref{MC_calculation_section}. In summary, we can conclude that the large $|q|$ modes contribute tiny to the mean value we want, but merely introduce large Monte Carlo error. Fortunately, Eq.~(\ref{Oxp2}) allows us to see a way to efficiently and accurately approximate $\langle \hat O \rangle$ since $S(\infty)=\nu(1-\nu)$ when $|q|\rightarrow\infty$ \cite{PrangeGirvin,haldaneselfdual}. Therefore we can introduce a cutoff $Q$, and separate the sum in Eq.~(\ref{Oxp2}) into short-ranged ($|q|>Q$) and long-ranged ($|q|<Q$) parts. Only for the long-ranged part, we calculate by Monte Carlo by using the lattice representation Eq.~(\ref{latticesum2}). For the short-ranged part, we simply replace $S(q)$ with $S(\infty)$ and calculate \emph{directly}.

Assuming $S(q)$ being saturated when $|q|>Q$ introduces systematic error $\delta E_S$. Although we don't know the short wavelength oscillation behavior of $S(q)$, we are still able to give an upper bound of $|\delta E_S|$, which could be calculated analytically. Note that $S(q)$ is positive and is bounded from above by its maximum value $S_{max}$, the oscillation must be less than $\min\{S^{max}-S(\infty), S(\infty)\}$. Hence, an upper bound of the systematic error is given by the following,
\beqn
|\delta E_S|&<&\frac{1}{4\pi N_{\phi}}\sum'_{|q|>Q} |O^{GC}(ql_B)|\cdot\delta S\label{systematicerror}\\
\delta S&=&\min\{S_{max} - S(\infty), S(\infty)\}.\nonumber
\eeqn
From the plot of the structure factor Fig~(\ref{lausq}) and Fig~(\ref{cflsq}), we empirically set $S_{max}\approx0.5$ for laughlin $\nu=1/3$ state, $S_{max}\approx0.8$ for CFL $\nu=1/2$ state.

This systematic error $\delta E_S$ must be included, together with the Monte Carlo error $\delta E_M$, into the total uncertainty $\delta E_{tot}=\sqrt{\delta E_M^2 + \delta E_S^2}$. Increasing the cutoff $Ql_B$ decreases $\delta E_S$ but makes $\delta E_M$ larger. The best value of $Ql_B$ is suppose to the one whose $\delta E_M$ and $\delta E_S$ are on the same order.

Table \ref{table:energy2} uses this approach to recalculate the Coulomb energies which were originally calculated in Table~\ref{table:energy}. We can see that by cutting off and approximating the large $q$ contribution we can significantly decrease the statistical error, and obtain improved estimates for the energy. 
\begin{table}
\begin{tabular}{|c|c|c|c|c|c|}
\hline
\multicolumn{6}{|c|}{$n=2$, $Ql_B=5$} \\
\hline
 $N_e$ & Exact & MC energy & $\delta E_M$ & $\delta E_S$ & $\delta E_{tot}$ \\
\hline
$4$ & $-0.280537$ & $-0.278$ & $0.004$ & 4e-4 & 0.004  \\
$5$ & $-0.278052$ & $-0.278$ & $0.005$ & 6e-4 & 0.005  \\
$6$ & $-0.274849$ & $-0.269$ & $0.005$ & 0.001 & 0.005 \\ 
$11$& $-0.268481$ & $-0.271$ & $0.006$ & 0.001 & 0.006 \\
$12$& $−0.268005$ & $-0.262$ & $0.006$ & 0.001 & 0.006 \\
\hline
\end{tabular}
\captionsetup{justification=RaggedRight, singlelinecheck=true}
\caption{By separating the energy calculation into short- and long-ranged parts and approximating the long-ranged part, we can dramatically reduce our statistical error. In this table we compute the Coulomb energy of the Laughlin wavefunction in $n=2$ and $n=3$. We used  the same number of Monte Carlo steps as we did in Table ~\ref{table:energy}, but find that our statistical error is reduced by up to two orders of magnitude even at these relatively small sizes. We also did test on larger systems, e.g. $N_e=11,12$.}
\label{table:energy2}
\end{table}

\subsection{Pair Amplitudes}
The self duality relation in Eq.~(\ref{self-dual}) implies the $S(q)$ can be expanded in terms of Laguerre polynomials (multiplied by Gaussians), which form a complete basis of polynomials that are self-dual under Fourier transformations. The expansion coefficients in this basis are known as `pair amplitudes'. Such pair amplitudes appear in the pseudopotential Hamiltonian, and therefore it is interesting to ask whether they can be calculated in our Monte Carlo method.

Before defining pair amplitude on torus, lets first look at the infinite plane geometry where the pair amplitude is better understood. The infinite plane has rotational symmetry, and angular momentum is well defined. A projector that projects a two-particle pair into a given relative momentum sector ($=m$) can be defined as,
\beqn
P^{ij}_m = 2\int \frac{d^2ql_B^2}{2\pi} L_m(q^2l_B^2)e^{-\frac{1}{2}q^2l_B^2} e^{iq(R_i - R_j)}.
\label{projector}
\eeqn
$P^{ij}_m$ are orthogonal projectors that satisfy
\beqn
P^{ij}_mP^{ij}_m &=& P^{ij}_m, \label{projector1} \\
P^{ij}_mP^{ij}_n &=& 0, \quad \text{if}\ m\neq n. \label{projector2}
\eeqn

The $m^{th}$ pair amplitude $\xi_m$ is the probability of finding particle pairs with relative angular momentum $=m$,
\beqn
\xi_m \equiv \langle \sum_{i<j}P_m^{ij} \rangle
\eeqn

On torus, $P^{ij}_m$ is defined similarly as in Eq.~(\ref{projector}), but with the integral over a continuum of momenta replaced by a discrete sum over all points in the reciprocal space. Since the torus does not have continuous rotation symmetry, the $m$ does not have the meaning of ``relative angular momentum'' any more, and $P^{ij}_m$ are no longer orthogonal: (\ref{projector2}) does not hold, while (\ref{projector1}) is modified
\beqn
P^{ij}_mP^{ij}_m &=& C_mP^{ij}_m. \label{torus_projector1}
\eeqn
where $C_m$ is a number that is slightly larger than one.
The fact (\ref{torus_projector1}) does not introduce any projectors with $m\neq n$ ensures that torus Laughlin wavefunction is still the exact ground state of the pseudo potential Hamiltonian. 
Although torus has only discrete rotation symmetry, the continuous rotation symmetry is restored and the $P^{ij}_m$ become orthogonal ($C_m\rightarrow1$) in the limit of $N_{\phi}\rightarrow\infty$.

The calculation of pair-amplitude shares the same spirit as high LL Coulomb energy. In the problem of calculating the pair-amplitude, we simply replace Eq.~(\ref{highcoulombogc}) with Eq.~(\ref{pairamplitdueogc}). Error analysis follows the same algorithm as discussed in the last section.
\beqn
O^{GC}(ql_B) = 4\pi [L_m(q^2l_B^2)e^{-\frac{1}{2}q^2l_B^2}]_c\label{pairamplitdueogc}
\eeqn
In Table~\ref{table:onethirdne6}, we calculated several orders of pair-amplitude for Laughlin $\nu=1/3$ state for $N_e=6$ particle.
\begin{table}
\centering
 \begin{tabular}{| c | c | c | c | c | c | c| } 
 \hline
 & ED & MC Value & $Ql_B$ & $\delta E_M$ & $\delta E_S$ & $\delta E_{tot}$  \\ [0.5ex] 
 \hline
  $\xi_1$ & 0.        & 1e-3 & $5.00$ & 1e-3 & 5e-4 & 1e-3 \\  
  $\xi_3$ & 5.928056  & 5.84 & $5.00$ & 0.08 & 0.04 & 0.09 \\ 
  $\xi_5$ & 4.441078  & 3.75 & $5.00$ & 0.8 & 0.7 & 1.08\\ 
 \hline
 \end{tabular}
 \captionsetup{justification=RaggedRight, singlelinecheck=true}
\caption{The pair-amplitude calculated for $N_e=6$ particles in Laughlin $\nu=1/3$ state. The $\delta E_{M}$ and $\delta E_S$ are Monte Carlo error and systematic error respectively. The total error $\delta E_{tot}=\sqrt{\delta E_M^2 + \delta E_S^2}$.}
\label{table:onethirdne6}
\end{table}

\section{Particle-Hole Overlap Through Monte-Carlo}
\label{sec:particlehole}

It is interesting to ask whether wavefunctions such as Eq.~(\ref{CFLwf}) are particle-hole symmetric. In \cite{scottjiehaldane}, we have addressed this question by numerically second-quantizing these wavefunctions, and then implementing particle hole symmetry in the second quantized basis by exchanging the filled and empty orbitals. Since we have now developed a tool for rapid calculations in the Schr\"odinger representation, it is natural to ask whether we can evaluate particle hole symmetry in this representation.

According to \cite{girvinph}, if we have some wavefunction $\Psi_1$, we can compute its particle-hole conjugate as follows:
\begin{equation}
\Psi_1^{PH}(z_j)=\int \prod_{i=1}^{N_e} dz_i ~~ \Psi_1(z_i) \Psi^*_{LL}(z_i,z_j)
\label{PH}
\end{equation}
where $\Psi_{LL}$ is the wavefunction for a filled Landau level. Using this definition of particle-hole conjugation we can compute the quantity $\langle \Psi^\beta_{CFL}(\{d\}) | PH \Psi^{\beta^\prime}_{CFL}(-\{d\}) \rangle$, which is the overlap between the CFL state and its particle-hole conjugate. Here $\beta$ indicates which center-of-mass sector the wavefunction is in, while $\{d\}$ represents the dipole moments of the wavefunction. Particle-hole symmetry on its own changes the momentun of a wavefunction, so when we write $PH$ we really mean particle-hole symmetry combined with a rotation by $\pi$, an operation which preserves the symmetry \cite{scottjiehaldane}. The $\pi$ rotation reverses the center-of-mass sector (so we will need $\beta \neq \beta^\prime$), and also takes $d\rightarrow -d$. Equivalently to reversing the $d$'s, we can instead reverse all the coordinates $z$, which is what we will do from now on. Using Eq.~(\ref{PH}), we can write the particle-hole overlap as follows:
\begin{eqnarray}
&&\langle \Psi^\beta_{CFL}(\{d\}) | PH \Psi^{\beta^\prime}_{CFL}(-\{d\}) \rangle=  \label{PHint}\\
&&\int \prod_{j=N_e+1}^{N_\Phi} dz_j \Psi^\beta_{CFL}(\{z_j\}) \times \nonumber\\ 
&&\int \prod_{i=1}^{N_e} dz_i ~~ \Psi_{CFL}^{\beta^\prime}(-\{z_i\}) \Psi^*_{LL}(\{z_i\},\{z_j\}) \nonumber\\
&&=\langle \psi_1 (x)|\psi_2(x)\rangle \nonumber\\
&&\psi_1(x)=\Psi^\beta_{CFL}(\{z_j\})\Psi_{CFL}^{\beta^\prime}(-\{z_i\})\label{PHpsi1}\\
&&\psi_2(x)=\Psi^*_{LL}(\{z_i\}, \{z_j\})\label{PHpsi2}
\end{eqnarray}
In the above equation we have stopped explicity writing the variational parameters $\{ d\}$, and as in Sec.~(\ref{MC_calculation_section}) we use $x$ as a shorthand for all the coordinates $\{z_i\}, \{z_j\}$.

In Sec.~(\ref{MC_calculation_section}), [specifically Eq.~(\ref{MCdef})] we were calculating the overlap $\langle \psi_1|\psi_2 \rangle$ and we manipulated the wavefunctions in such a way that $|\psi_1|^2$ could be used as a Metropolis weight.
 However, we could in principle use any real, non-negative function as a weight. 
 This inspires more general version of Eq.~(\ref{MCdef}):
 \begin{eqnarray}
\frac{\langle\psi_1|\psi_2\rangle}{\sqrt{\langle\psi_1|\psi_1\rangle\langle\psi_2|\psi_2\rangle}}=\frac{\sum' O_{12}(x) p(x) }{\sqrt{[\sum' O_{11}(x) p(x)][\sum' O_{22}(x) p(x)]}}.\nonumber\\\label{MCdef2}
\end{eqnarray} 
where $O_{ij}(x)\equiv \frac{\psi_i^*(x)\psi_j(x)}{p(x)}$. Here $p(x)$ is the statistical weight, so it must be real and non-negative. $|\psi_1|^2$ is a good choice for $p(x)$ when $\psi_1$ and $\psi_2$ are very similar, because it means that $O(x)$ will be order one, and this is necessary for efficient importance sampling. If $O(x)$ can vary widely then we will no longer be doing importance sampling (i.e. configurations where $O(x)p(x)$ is large will not be sampled frequently) and the algorithm will be inefficient. If $p(x)=0$ when $\psi_1(x)$ [or $\psi_2(x)$] is non-zero
 the Monte Carlo will give wrong results since $O_{11}(x)$ [or $O_{22}(x)$] is infinite.
  
The $\psi_1$ and $\psi_2$ defined in Eqs.~(\ref{PHpsi1}-\ref{PHpsi2}) are not very similar, and in fact one can have zeros where the other one is large. 
A simple way to see this is that whenever $z_i=z_j$ for any $i,j$ in Eq.~(\ref{PHint}), $\psi_2$ will vanish but $\psi_1$ does not have to. Therefore simply using $|\psi_1|^2$ or $|\psi_2|^2$ for $p(x)$ will not work. In this work we make the following choice for $p(x)$:
\beqn
p(x)=(\alpha|\psi_1(x)| + |\psi_2(x)|)^2.
\label{pz}
\eeqn 
The virtues of this choice is that $p(x)$ will be large whever either $\psi_1$ or $\psi_2$ is large. The CFL wavefunctions are not normalized so the parameter $\alpha$ is included to make the two terms in the sum of approximately equal size. Using a fixed value of $\alpha$ (e.g.~$\alpha=1$) will give correct results but tuning $\alpha$ for a given system size can dramatically improve the performance of the Monte Carlo. We find for the wavefunctions used in this paper that $|\psi_1(x)|^2$ is roughly two orders orger of Note that other choices of $p(x)$ are possible so long as it is large whenever either wavefunction is large, it maybe be possible to further improve performance with a better choice of $p(x)$.
 
A final obstacle to computing the particle-hole overlap is that the wavefunctions produced by Eq.~(\ref{PH}) are not normalized, even in the wavefunctions on the right-hand side of that equation are normalized. In order to obtain a normalized wavefunction (and therefore a sensible overlap) we need to multiply Eq.~(\ref{PH}) and (\ref{PHint}) by a normalization constant $\sqrt{C}$, where 
\begin{equation}
C=\left(\begin{array}{c} N_\Phi \\ N_e \end{array} \right).
\end{equation} 
The value of this constant can be explained by thinking about the overlap we are calculating as an overlap of the wavefunctions $\psi_1$, $\psi_2$ defined in Eqs.~(\ref{PHpsi1}-\ref{PHpsi2}). If the two CFL wavefunctions were particle-hole symmetric, this overlap would be $1$. But wavefunction $\psi_1$ is completely antisymmetric under interchanging coordinates $z_i$ (which appears in one CFL wavefunction) and $z_j$ (which appears in the other wavefunction). In order for the overlap to be $1$, $\psi_2$ must therefore also have this symmetry, but it clearly does not. Therefore to get sensible results we must antisymmetrize Eq.~(\ref{PHpsi2}). Each term in such an antisymmetrization will be exactly the same once all positions are summed over, but in order to stay normalized we must divide by the square root of the number of terms in the antisymmetrization, which is exactly $\sqrt{C}$.

\begin{figure}
\includegraphics[width=0.8\linewidth]{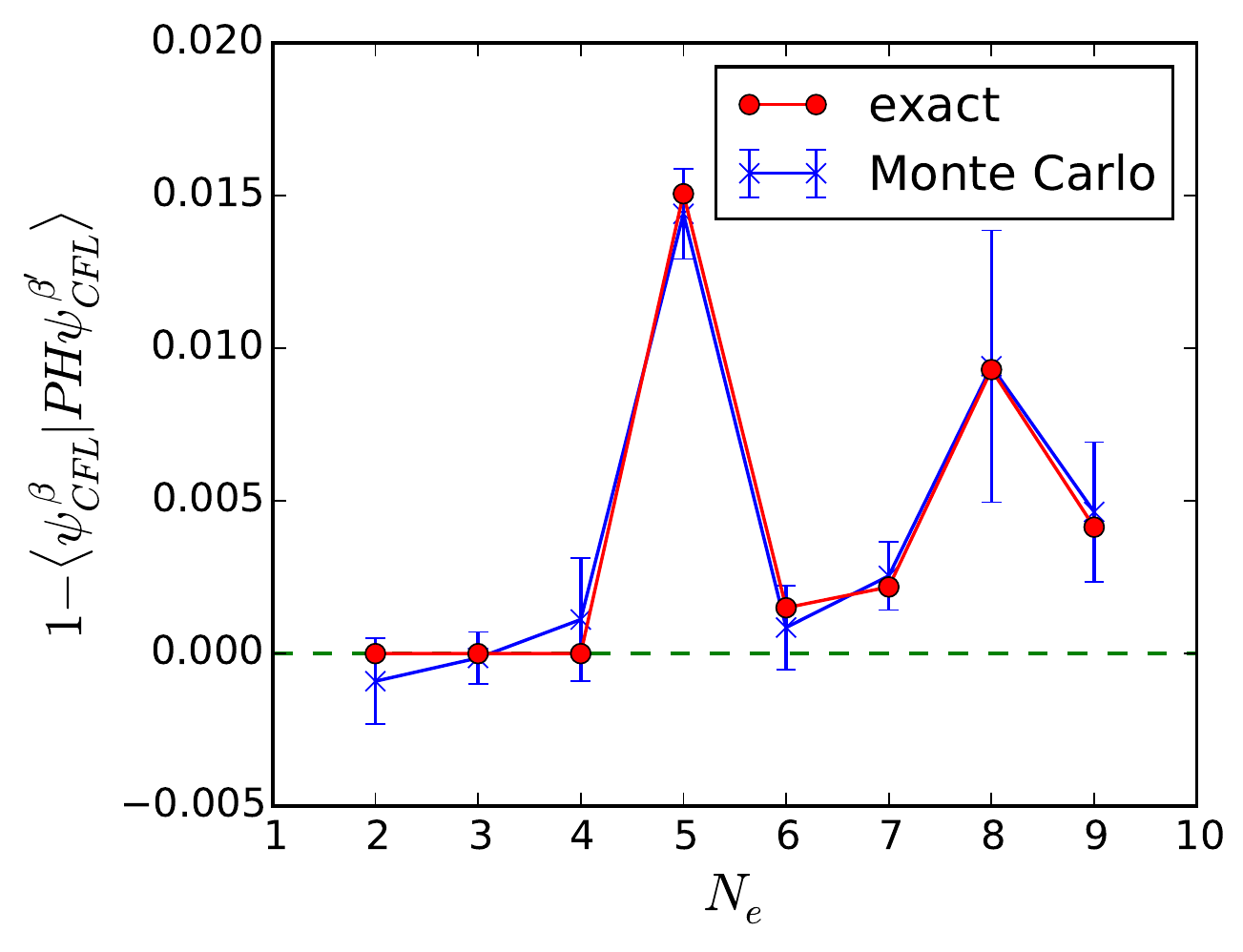}
\caption{ Overlaps between a wavefunction and its PH conjugate. The red points from come from doing an exact second-quantization of the model wavefunction as in Ref.~\onlinecite{scottjiehaldane}, while the blue comes from a Monte Carlo calculation. For each $N_e$, the configuration of $d$'s with the largest overlap was used.}
\label{PHfig}
\end{figure}

This normalization constant means both the values produced by numerically computing Eq.~(\ref{MCdef2}) {\it and their statistical errors} must be multiplied by $\sqrt{C}$. Therefore to keep the statistical errors constant in system size, the number of Monte Carlo steps requires scales as $\propto C$. This is the same algorithmic complexity as numerically second-quantizing the wavefunction, as in Ref.~\onlinecite{scottjiehaldane}. Therefore there is no benefit to using our Monte Carlo method to compute particle-hole overlaps. Nevertheless the algorithm does work, as can be seen in Fig.~\ref{PHfig} where we show the particle-hole overlaps for a few values of $N_e$, and compare them to the results of numerical second quantization. The $N_e=9$ data in Fig.~\ref{PHfig} took ~600 CPU hours, while doing the exact second-quantizing algorithm takes around ten minutes. Therefore though using Monte Carlo does give correct results it is not a practical method to evaluate the particle-hole symmetry of model wavefunctions.

\section{Discussion}
We have shown that for quantum Hall problems on a torus in a single Landau level, continuum integrals can be replaced by sums over a lattice of spacing $L/N_{\phi}$. This procedure can be used to dramatically save the time required for Monte Carlo calculations, because the continuous sampling is redundant and the special functions required for quantum Hall wavefunctions on a torus can be tabulated in advance. We used our procedure to calculate a number of quantities, such as the energy of model wavefunctions, quasiparticle braiding statistics, the Berry phase acquired by composite fermions moving around the composite Fermi surface, guiding center structure factors and the particle-hole symmetry of model wavefunctions.

Our method can be used to dramatically increase the accessible system sizes for almost every quantity calculated using Monte Carlo. There are a few quantities which we still do not know how to calculate, for example the real-space entanglement entropy, applying our formalism to such methods is an interesting direction for future work.

\acknowledgments
This work was supported by Department of Energy BES Grant DE-SC0002140.


\appendix
\section*{Appendix A: Modified Weierstrass Sigma Function}
The torus wavefunctions used in the main text all rely on the function we call $f(z)$. In this section, we will explain how this function is constructed. 
The definition of $f(z)$ is:
\beqn
f(z) &=& \sigma(z)e^{-\frac{1}{2N_{\phi}}zz^*}. \label{f_def}\\
\sigma(z) &=& \tilde\sigma(z)e^{-\frac{1}{2}\bar G(\mathbb L) z^2}. \label{defofsigma}
\eeqn
i.e. $f(z)$ is a Gaussian factor multiplied by a holomorphic function $\sigma(z)$, which we call the ``modified sigma function''. 
The modified sigma function is designed by multiplying the standard Weierstrass sigma function $\tilde\sigma(z)$ and a holomorphic factor $e^{-\frac{1}{2}\bar G(\mathbb L) z^2}$. The $\bar G_2(\mathbb L)$ is a modular independent $c-$number constant that vanishes for square and hexagonal torus. Note that $f(z)$ is modular invariant, this is one of the key advantages to using it over the previously used Jacobi theta function \cite{haldanetorus1}.

The standard Weierstrass sigma function $\sigma(z)$ has a product series expansion,
\beqn
\tilde\sigma(z) \equiv z\prod_{L\in L_{mn}\backslash\{0\}}\left(1-\frac{z}{L}\right)e^{\frac{z}{L}+\frac{1}{2}\frac{z^{2}}{L^{2}}} \label{prod}
\eeqn
where $L_{mn} = \{mL_1 + nL_2\}$ defines the 2D torus. Clearly, it is modular invariant. It is also quasi-periodic,
\beqn
\tilde\sigma(z+L_i) = - e^{2\tilde\eta_i(z+L_i/2)}\tilde\sigma(z),\label{modsigmatrans}
\eeqn
where $\tilde\eta_i$ is the standard zeta function evaluated at half period, which is related to the $k=1$ Eisenstein series $G_2(L_i)$, $i=1,2$,
\beqn
\tilde\eta_i &=& G_2(L_i)L_i/2.\label{eta-def}
\eeqn
The Eisenstein series $G_2(L_i)$ has a highly convergent, \emph{numerical feasible} formula,
\beqn
G_2(L_i) &=& \frac{2\pi^2}{L_i^2}\left(\frac 1 6 + \sum_{n=1}^{\infty}\frac{1}{\sin^2(n\pi\frac{L_{j\neq i}}{L_i})}\right),
\eeqn
The $\tilde\eta_i$ in addition obey a relation that defines chirality,
\beqn
\tilde\eta_1L_2 - \tilde\eta_2 L_1 = \frac{L_1^*L_2 - L_1L_2^*}{2N_{\phi}} = i \pi. \label{chirality}
\eeqn
The (\ref{eta-def}) and (\ref{chirality}) suggests a new modular independent quantity, called the ``almost modular form'',
\beqn
\bar G(\mathbb L) \equiv G_2(L_i) - \frac{1}{N_{\phi}}\frac{L_i^*}{L_i} \label{almostmf}
\eeqn
Now we define the modified sigma function $\sigma(z)$ through Eq.~(\ref{defofsigma}). It is easy to verify that,
\beqn
\sigma(z+L_i) = - e^{\frac{L_i^*}{N_{\phi}}(z+L_i/2)}\sigma(z).
\eeqn
This is the desired translation property. Combining the above equation with Eq.~(\ref{f_def}) gives us the translation properties of $f(z)$,
\beqn
f(z+L) &=& \eta_L f(z)e^{\frac{1}{2N_{\phi}}(L^*z-Lz^*)}.
\label{ftrans}
\eeqn
where $\eta_L = \pm 1$ [note it's not the zeta function $\tilde\eta$], which $=1$ if $\frac{1}{2}L\in\mathbb{L}$, and $-1$ otherwise.

\section*{Appendix B: Expression of Two-Body Lattice Operator}
In this section, we derive the expression in Eq.~(\ref{latticesum2}). We will start by discussing the one-body operator. As will seen later two-body operator is a simple generalization.\\
\\
We will ask the question, what is the lattice summation value of a single-body operator $\langle\psi_1|\hat O^{Lat}|\psi_2\rangle_{Lat}$, where $|\psi_{i=1,2}\rangle$ being the lowest Landau level single-body states. To find this, first notice that,
\beqn
\int d^2x\ \psi_1^*(x)\psi_2(x)e^{iqx} = f_0(q)\langle\psi_1^{GC}|e^{iqR}|\psi_2^{GC}\rangle
\eeqn
Doing the inverse Fourier transformation for the above equation, we arrive at,
\beqn
\psi_1^*(x)\psi_2(x) = \frac{1}{(2\pi)^2}\sum_q f_0(q)e^{-iqx}\langle\psi_1^{GC}|e^{iqR}|\psi_2^{GC}\rangle.\nonumber
\eeqn
Since we use the periodic boundary condition, $e^{iq_{mn}N_{\phi}R}|\psi^{GC}\rangle = (-1)^{N_{\phi}(nm+m+n)}|\psi^{GC}\rangle$. The above equation, after compactifying into the first Brillouin zone, becomes
\beqn
\psi^*_1(x)\psi_2(x) = \frac{1}{(2\pi)^2}\sum'_q[f_0(q)]_{N_{\phi}}e^{-iqx}\langle\psi_1^{GC}|e^{iqR}|\psi_2^{GC}\rangle.\nonumber
\eeqn
where,
\beqn
[f_0(q_{mn}l_B)]_{N_\phi} &\equiv& \sum_{q'} f_0(q_{mn}l_B + q'_{kl}N_{\phi}l_B)\label{f0nphi}\\
&\times& (-1)^{ml-nk+N_{\phi}(kl+k+l)}.\nonumber
\eeqn
By doing a discrete real space summation, we find and conclude that, for any single body operator defined on lattice $O(x) = \frac{1}{2\pi N_{\phi}}\sum'_{q}O(q)e^{iqx}$, there is,
\beqn
& & \sum'_x\psi_1^*(x)\psi_2(x)O(x)\label{comp2}\\
&=& \frac{1}{2\pi N_{\phi}}\sum'_q O(q)[f_0(q)]_{N_{\phi}}\langle\psi_1^{GC}|e^{iqR}|\psi_2^{GC}\rangle.\nonumber
\eeqn
This is the lattice representation for single body operators. Now, it's ready to prove the central results used in main context Eq.~(\ref{latticesum2}) for two-body operators:
\beqn
&&  \frac{1}{2\pi N_{\phi}}\sum_q\sum'_{x_i}\psi_1^*(x_1)\psi_2^*(x_2)O(q)e^{iq(x_1-x_2)}\psi_3(x_2)\psi_4(x_1)\nonumber\\
&=& \frac{1}{2\pi N_{\phi}}\sum_q\sum'_{x_i}\psi_1^*(x_1)\psi_4(x_1)e^{iqx_1}[\psi_3^*(x_2)\psi_2(x_2)O(q)e^{iqx_2}]^*\nonumber
\eeqn
where $O(x_1-x_2) = \frac{1}{2\pi N_{\phi}}\sum_qO(q)e^{iq(x_1-x_2)}$ is the two-body operator, the $O(q)$ is thus periodic in $q$. Applying Eq.~(\ref{comp2}) twice gives,
\beqn
&&  \frac{1}{2\pi N_{\phi}}\sum_q\sum'_{x_i}\psi_1^*(x_1)\psi_2^*(x_2)O(q)e^{iq(x_1-x_2)}\psi_3(x_2)\psi_4(x_1)\nonumber\\
&=& \frac{1}{2\pi N_{\phi}}\sum'_qO(q)|[f_0(q)]_{N_{\phi}}|^2\label{twobodylat}
\eeqn
By comparing the above with the guiding center space interactions Eq.~(\ref{latop}), we conclude that the lattice representation for two-body operators is Eq.~(\ref{latticesum2}).

\bibliography{mcpaper.bib}
\end{document}